\documentclass{llncs}
\usepackage{ifpdf}
\usepackage{makeidx}  
\usepackage{graphics,graphicx}
\begin{document}
\mainmatter              
\newcommand{\pn}{\mathrm{Min}}
\newcommand{\px}{\mathrm{Max}}
\newcommand{\Val}{\mathrm{Val}}
\newcommand{\Width}{\mathrm{Width}}
\ifpdf
\DeclareGraphicsRule{*}{mps}{*}{}
\else
\DeclareGraphicsRule{*}{eps}{*}{}
\fi
\title{Simple Recursive Games}
\titlerunning{Simple Recursive Games}  
%
\author{Daniel Andersson \and Kristoffer Arnsfelt
  Hansen \and\\ Peter Bro Miltersen \and Troels Bjerre S\o{}rensen}
\authorrunning{Daniel Andersson et al.}   
\tocauthor{Daniel Andersson, Kristoffer Arnsfelt
  Hansen, Peter Bro Miltersen and Troels Bjerre S\o{}rensen (University
  of Aarhus)}
\institute{Department of Computer Science, University of Aarhus, Denmark\\
\email{\{koda,arnsfelt,bromille,trold\}@daimi.au.dk}}

\maketitle              

\begin{abstract}
We define the class of {\em simple recursive games}. A simple
recursive game is defined as a simple stochastic game (a notion
due to Anne Condon), except that we allow arbitrary real payoffs but disallow moves of chance. We study the complexity
of solving simple recursive games and obtain an almost-linear time
comparison-based algorithm for computing an equilibrium of such a game. 
The existence of a linear time comparison-based algorithm remains 
an open problem. 
%
\end{abstract}
\section{Introduction}
\label{intro}
Understanding rational behavior in {\em infinite duration games}
has been an important theme in pure as well as computational game
theory for several decades. A number of central problems remain
unsolved. In pure game theory, the existence
of near-equilibria in general-sum two-player stochastic games
were established in a celebrated result by Vieille \cite{Vieille:2000:I,Vieille:2000:II}, 
but the existence of
near-equilibria for the three-player case remains an important
and elusive open problem \cite{Aumann:2003:Presidential}. In computational
game theory, Condon \cite{Condon:1992:CSG} delineated the efficient computation of 
positional equilibria in {\em simple stochastic games} as an important
task. While Condon showed this task to be doable
in ${\mathbf{NP}} \cap {\mathbf{coNP}}$, to this day, the best deterministic algorithms
are not known to be of subexponential complexity. To the computer
science community, the problem of computing
positional equilibria in simple stochastic games is motivated by 
its hardness for finding equilibria in many other natural classes
of games \cite{Zwick:1996:CMPG}, which again implies hardness for tasks
such as model checking the $\mu$-calculus \cite{Emerson:2001:MCMC}, 
which is relevant for the {\em
formal verification} of computerized systems.

\subsection{Simple Stochastic Games}
A simple stochastic game \cite{Condon:1992:CSG} is given by a graph 
$G=(V,E)$. The vertices in $V$ are the {\em positions} of the game.
Each vertex belongs either to player {\em Max}, to
player {\em Min}, or to {\em Chance}. There is a 
distinguished \emph{starting
position} $v_0$. Furthermore, there are a number of distinguished {\em
  terminal positions} or just {\em terminals}, each labeled with a
{\em payoff} from Min to Max.\footnote{In Condon's original paper, there were only two terminals,
with the payoffs $0$ and  $1$. 
The relaxation to
arbitrary payoffs that we adopt here is fairly standard.
} All positions except
the terminal ones have outgoing arcs. The game is played by
initially placing a token on $v_0$, letting the token move along
a uniformly randomly chosen outgoing arc when it is in a position belonging to Chance
and letting each of the players decide along which outgoing arc to move the token
when it is in a position belonging to him. If a
terminal is reached, then Min pays Max its payoff and the game
ends.
Infinite play yields payoff $0$. A {\em positional
strategy} for a player is a selection of one outgoing arc for each of
his positions. He plays according to the strategy
if he moves along these arcs whenever he is to move.  
It is known (see \cite{Condon:1992:CSG}) that each position $p$ in a simple 
stochastic game can be assigned a {\em value} $\Val(p)$ so that:
\begin{enumerate}
\item{}\label{prop1}Max has a positional strategy that, regardless of what
  strategy Min adopts, ensures an expected payoff of at least
  $\Val(p)$ if the game starts in $p$.
\item{}\label{prop2}Min has a positional strategy that, regardless of what
strategy Max adopts, ensures that the expected payoff
is at most $\Val(p)$ if the game starts
in $p$.
\end{enumerate}
The value of the game itself is the value of $v_0$.
Condon considered the complexity of computing this value. It
is still open if this can be done in polynomial time. In the
present paper, we shall look at some easier problems. For
those, we want to make some distinctions which are inconsequential
when considering whether the problems are polynomial time solvable
or not, but important for the more precise (almost linear) time 
bounds that we will be interested in in this paper.
\begin{itemize}
\item{}A {\em weak solution} is $\Val(v_0)$
{\em and} a positional strategy for each player satisfying the
conditions in items \ref{prop1} and
\ref{prop2} above for $p=v_0$.
\item{}
A {\em strong solution} is the list of values
of {\em all} positions in the game {\em and} a positional strategy for
each player that for {\em all} positions $p$, ensures an
expected payoff of at least/most $\Val(p)$ if the game starts in $p$.
\end{itemize}
In game theory terminology, weak solutions to a game are 
{\em Nash equilibria} of the game while strong solutions are 
{\em subgame perfect equilibria}. 
Figure~\ref{fig:subgame} illustrates that the distinction is not
inconsequential. 
\begin{figure}[!h]
\begin{center}
\includegraphics{fig.1}\includegraphics{fig.2}
\end{center}
\caption{\label{fig:subgame}The left solution is weak, the right is strong.}
\end{figure}
In the weak solution to the left, Max (if he gets to move) is content 
to achieve payoff $0$, the value
of the game, even though he could achieve payoff $1$.
Note that the game in Figure~\ref{fig:subgame} is {\em acyclic}. 
In contrast to the general case, it is of 
course well known that a strong solution to an acyclic game can be 
found in 
linear time by straightforward dynamic
programming (known as {\em backwards induction} in the game theory community).
We shall say that we weakly (resp. strongly) solve a given game when
we provide a weak (resp.
strong) solution to the game.
Note that when talking about strong solutions, the starting position is irrelevant and does not
have to be specified.

\subsection{Simple Recursive Games}
\label{sorting_method_section}
Condon established that for the case of a simple stochastic game
with {\em no Chance positions} and only 0/1 payoffs, the game can be strongly
solved in {\em linear time}. 
Interestingly, Condon's algorithm has been discovered and described
independently by the artificial intelligence community where
it is known under the name of {\em retrograde analysis} 
\cite{Thompson:1986:RACE}. It is
used routinely in practice for finding optimal strategies for 
combinatorial games 
that are small enough for the game graph to be represented in (internal
or external) memory and where dealing with
the possibility of cycling is a non-trivial aspect of the
optimal strategies. The best known example is the construction of
tables
for {\em chess endgames} \cite{Heinz:1999:SSCC}.
Condon's algorithm (and retrograde analysis) being linear time depends
crucially on the fact that the games considered are 
win/lose games (or, as is usually the case in the AI literature, win/lose/draw
games), i.e., that terminal payoffs are either $0$ or $1$ (or 
possibly also $-1$, or in some AI examples even a small range of 
integers, e.g., \cite{Romein:2003:SGA}).
In this paper we consider the algorithmic problem arising when
{\em arbitrary real payoffs} are allowed. That is, we consider a class
of games similar to but incomparable to Condon's simple stochastic games: 
We disallow chance vertices, but allow arbitrary real payoffs. We call
the resulting class {\em simple recursive games}.\footnote{
The perfect information, no chance special case of Everett's
``recursive games'' \cite{Everett:1957:RG}.
}

Some simple examples of simple recursive games are given in Figure~\ref{cycle_games}.
In (a), the unique strong solution
is for Min to choose right and for Max to choose left.
Thus, the outcome is infinite play.
\begin{figure}
\begin{center}
$\begin{array}{c@{\hspace{1in}}c}
\includegraphics{fig.4} &
\includegraphics{fig.5} \\ [0.4cm]
\mbox{\bf (a)} & \mbox{\bf (b)}
\end{array}$
\end{center}
\caption{{\bf (a)} Infinite play equilibrium. {\bf (b)} All values are $1$,
  but one choice is suboptimal.}
\label{cycle_games}
\end{figure}
In (b), the unique strong solution
is for Min to choose right and for Max to choose right.
The values
of both vertices are $1$, but we observe that it is {\em not}
a sufficient criterion for correct play to choose a vertex
with at least as good a value as your current vertex. In
particular, according to this criterion, Max could choose
left, but this would lead to infinite play and a payoff of
$0$, which is a suboptimal outcome for Max.

We observe below that if a {\em sorted} list of the
payoffs (with pointers to the corresponding terminals of the
game) is given in advance, optimal strategies can again be found
in linear time without further processing of the payoffs. 
From this it follows that a simple recursive game 
with $n$ payoffs and $m$ arcs in the graph can be solved in time 
$O(n \log n + m)$ by a {\em comparison-based} algorithm.
 The main question we attempt to approach in this paper is the following:
\vspace{1em}

\noindent
{\bf Main Open Problem.}
{\em Can a simple recursive game be (weakly or strongly) solved 
in linear time by a
comparison-based algorithm?} 
\vspace{1em}

\noindent We believe this to be an interesting question, both in the context
of game solving (simple recursive games being a very simple yet non-trivial
natural variant of the general problem)
and in the context of the study of comparison-based algorithms
and comparison complexity.
This paper provides neither a positive nor a negative answer to the
question, but we obtain a number of partial results, described in the
next subsection.
 
\subsection{Our Results}
Throughout this section we consider simple recursive games with 
$n$ denoting the number of terminals  (i.e., number of
payoffs) and $m$ denoting the total size (i.e., number of arcs) 
of the graph defining the game. 
We can assume $m \geq n$, as terminals
without incoming arcs are irrelevant.
\vspace{-.5em}
\subsubsection*{Strategy Recovery in Linear Time.} The example of
Figure \ref{cycle_games} (b) shows that it is not completely trivial to
obtain a strong solution from a list of values of the vertices.
We show that this task can be done in linear time, i.e. time $O(m)$.
Thus, when constructing algorithms for obtaining a strong solution,
one can concentrate on the task of computing the values $\Val(p)$ for
all $p$. Similarly,
we show that given the value of just the starting position, a
weak solution to the game can be computed in linear time. 
\vspace{-.5em}
\subsubsection*{The Number of Comparisons.}
When considering comparison-based algorithms, it is natural to
study the number of comparisons used separately from the running time 
of the algorithm (assuming a standard random access machine). By
an easy reduction from sorting, we show that there is no comparison-based
algorithm that {\em strongly} solves a given game using only $O(n)$ 
comparisons. In fact, $\Omega(n \log n)$ comparisons are necessary.
In contrast, Mike Paterson (personal communication) has
observed that a simple recursive game can be {\em weakly} solved using $O(n)$
comparisons and $O(m \log n)$ time. With his kind permission, his
algorithm is included in this paper. This also means that for the case of
weak solutions, our main open problem cannot
be solved in the negative using current lower-bound techniques, as
it is not the number of comparisons that is the bottleneck.
Our lower bound uses a game with $m = \Theta(n \log n)$ arcs. Thus,
the following interesting open question concerning only the
comparison complexity remains:
{\em Can a simple recursive game be {\em strongly} solved using $O(m)$
  comparisons?} If resolved in the negative, it will resolve
our main open problem for the case of strong solutions.
\vspace{-.5em}
\subsubsection*{Almost-Linear Time Algorithm for Weak Solutions.}
As stated above, Mike Paterson has 
observed that a simple recursive game can be weakly
solved using $O(n)$ comparisons and $O(m \log n)$ time. We refine his
algorithm and obtain an algorithm that weakly solves a game
using $O(n)$ comparisons and only
$O(m \log \log n)$ time. Also, we obtain an algorithm that weakly 
solves a game in time $O(m + m (\log^* m -\log^* \frac{m}{n}))$ 
but uses a superlinear number of comparisons. 
For the case of {\em strongly} solving a game,
we have no better bounds than those derived from the simple algorithm
described in Section \ref{sorting_method_section}, i.e., $O(m + n \log
n)$ time and $O(n \log n)$ comparisons.
Note that the bound
$O(m + m (\log^* m -\log^* \frac{m}{n}))$ is linear in $m$ whenever
$m \geq n \log \log \ldots \log n$ for a constant number of 'log's.
Hence it is at least as good a bound as $O(m + n \log n)$,
for any setting of the parameters $m,n$.
\section{Preliminaries}
\label{prelim}
\begin{definition}
A \emph{simple recursive game (SRG)} is a digraph with vertices
partitioned into sets of non-terminals $V_{\pn}$ and $V_{\px}$, which
are game positions where player $\pn$ and $\px$, respectively,
chooses the next move (arc), and terminals $T$, where the game
ends and $\pn$ pays $\px$ the amount specified by $p : T \to
\mathbf{R}$.\qed
\end{definition}

For simplicity, we will assume that terminals have distinct payoffs, i.e., that $p$ is
injective. We can easily simulate this by artificially distinguishing
terminals with equal payoffs
in some arbitrary (but consistent) fashion. We will also assume that
$m\geq n$, since terminals without incoming arcs are irrelevant.

\begin{definition}
We denote by $\Val_G(v)$ the value of the game $G$ when the vertex $v$ is
used as the initial position and infinite play is interpreted as a
zero payoff. This will also be called ``the value of $v$ (in $G$)''.
\qed
\end{definition}
\begin{remark}
That such a value indeed exists will follow from Proposition \ref{sorting_method}.  We shall later see how to construct optimal \emph{strategies} from vertex values.
\end{remark}
\begin{definition}
To \emph{merge} a non-terminal $v$ with a terminal $t$ is to remove
all outgoing arcs of $v$, reconnect all its incoming arcs to $t$, and
then remove $v$.
\qed
\end{definition}
The definitions of a {\em strong}  and a {\em weak} solution 
are as stated in the introduction.
The following algorithm is a generalization of Condon's linear time
algorithm 
\cite{Condon:1992:CSG} 
for solving simple recursive games with payoffs in $\{0,1\}$.
That algorithm is known as \emph{retrograde
  analysis} in the AI community \cite{Thompson:1986:RACE}, and we
shall adopt this name also for this more general version.
\begin{proposition}
\label{sorting_method}
Given an SRG and a permutation that orders its terminals, we can
find a strong solution to the game in linear time and using no further
comparisons of payoffs.
\end{proposition}
\begin{proof}
If all payoffs are 0, then all values are $0$ and
every strategy is optimal.

Suppose that the minimum payoff $p(t)$ is negative.
Any incoming arc to $t$ from a
$\px$-vertex that is not the only outgoing arc from that vertex is
clearly suboptimal and can be discarded. Each other incoming arc is an \emph{optimal
 choice} for its source vertex, which can therefore be merged with
 $t$. Symmetric reasoning applies when the maximum payoff is positive.
\qed\end{proof}
This immediately yields the \emph{sorting method} for strongly solving
SRGs: First sort the payoffs, and then apply Proposition \ref{sorting_method}.
\begin{corollary}
An SRG with $m$ arcs and $n$ terminals can be strongly solved in $O(m + n\log n)$ time.\qed
\end{corollary}

\begin{definition}
To \emph{merge} a terminal $s$ with another terminal $t$ is to
reconnect all incoming arcs of $s$ to $t$ and then remove $s$.
Two terminals are \emph{adjacent} if their payoffs have
the same sign and 
no other terminal has a payoff in between.
\qed
\end{definition}
The following lemma states the intuitive fact that when we merge two
adjacent terminals, the only non-terminals affected are those with the
corresponding values, and they acquire the same (merged) value. 
\begin{lemma}\label{lemma1}
If $G'$ is obtained from the SRG $G$ by merging a terminal $s$
with an adjacent terminal $t$, then for each non-terminal $v$, we have
\begin{equation}
\Val_{G'}(v) = \left \{ \begin{array}{ll} \Val_G(v) &\quad\mbox{if
      $\Val_G(v) \neq \Val_G(s)$,} \\
\Val_{G'}(t) &\quad\mbox{if
      $\Val_G(v) = \Val_G(s)$.} \end{array} \right.
\end{equation}
\qed\end{lemma}
\begin{proof}
Consider all SRGs with a fixed structure (i.e., underlying graph)
but with varying payoffs. Since a strong solution can 
be computed by a
comparison-based algorithm (Proposition \ref{sorting_method}), 
the value of any particular position $v$
can be described by a min/max formula over the payoffs. The 
claim of the lemma can be seen to be true
by a simple induction in the size of the relevant formula.
\end{proof}

By repeatedly merging adjacent terminals, we ``coarsen'' the
game. Figure~\ref{coarsening} shows an example of this. The
partitioning method we shall use to construct coarse games in this
paper also yields sorted lists of their payoffs. Hence, we shall be
able to apply retrograde analysis to solve them in linear time.
\begin{figure}
\begin{center}
$\begin{array}{c@{\hspace{1in}}c}
\includegraphics{fig.6} &
\includegraphics{fig.7} \\ [0.4cm]
\end{array}$
\end{center}
\caption{Coarsening by merging $\{-4,-1\}$ and $\{2,3,5\}$.}
\label{coarsening}
\end{figure}
\begin{corollary} \label{sgn}
The signs of the values of all vertices in a given SRG can be determined in linear time.
\end{corollary}
\begin{proof}
Merge all terminals with negative payoffs into one, do likewise for
those with positive payoffs, 
and then solve the
resulting coarse ``win/lose/draw'' game by retrograde analysis.
\qed\end{proof}
Clearly, arcs between vertices with different values cannot be part
of a strong solution to a game. 
From this, the following lemma is
immediate. 
\begin{lemma}
\label{splitup}
In an SRG, removing an arc between two vertices with different values does not
affect the value of any vertex.
\qed
\end{lemma}
\begin{remark}
Corollary \ref{sgn}, Lemma \ref{splitup}, and symmetry together allow us to restrict our attention to games where all vertices
have positive values, as will be done in subsequent sections.
\end{remark}
\begin{proposition}
Given the value of the initial position of an SRG, a weak solution
can be found in linear time. If the values of \emph{all}
positions are known, a strong solution can be found in linear time.
\end{proposition}
\begin{proof}
In the first case, let
$y$ be the value of initial position $v_0$.
We partition payoffs in at most five intervals:
$(-\infty, \min(y,0))$, $\{\min(y,0)\}$, $(\min(y,0), \max(y,0))$,
$\{\max(y,0)\}$ and $(\max(y,0),
\infty)$.
%
We merge all terminals in each of the intervals, obtaining a
game with at most five terminals.
A strong solution for the resulting coarse game is found in
linear time by
retrograde analysis. The pair of strategies obtained is then
a weak solution to the original game, by Lemma \ref{lemma1}. 

In the second case, by Lemma \ref{splitup},
we can first discard all arcs between vertices
of different values. This
disintegrates the game into
smaller games where all
vertices have the same value. We find a strong solution to each of
these games in linear time using
retrograde analysis. Combining these solutions in 
the obvious way
yields a strong solution to the original game, by Lemma \ref{splitup}.
\qed\end{proof}


\section{Solving Simple Recursive Games}
\subsection{Strongly}
\label{tradeoff}
For solving SRGs in the strong sense, we currently know no
asymptotically faster method than completely sorting the
payoffs. Also, the number of \emph{comparisons} this method performs is,
when we consider bounds \emph{only depending on the number of
  terminals} $n$, optimal. Any
sorting network \cite{Knuth:1997} can be
implemented by an acyclic SRG, by simulating each comparator by a Max-vertex and
 a Min-vertex. Figure \ref{fig:sorting} shows an example of
 this. Thus, we have the following tight bound.
\begin{proposition}
Strongly solving an SRG with $n$ terminals requires $\Theta(n\log n)$
comparisons in the worst case.
\qed
\end{proposition}
\begin{figure}
\begin{center}
\includegraphics{fig.3}
\end{center}
\caption{\label{fig:sorting}Implementing a sorting network by a simple
recursive game.}
\end{figure}
Implementing the
asymptotically optimal AKS-network \cite{AjtaiKomlosSzemeredi:1983} results in a game with $\Theta(n \log n)$
vertices and arcs. Thus, it is still consistent with our current
knowledge that a game can be strongly solved using $O(m)$ comparisons. 
\subsection{Weakly}
The algorithms we propose for weakly solving SRGs all combine
coarsening of the set of payoffs with
retrograde analysis. By splitting the work between
these two operations in different ways, we get different time/comparison trade-offs. At one
extreme is the sorting method. At the other, we partition the
payoffs around their median (which can be done in
linear time 
by Blum {\em et al.} \cite{BlumFloydPrattRivestTarjan:1972}),
use retrograde analysis to solve the coarse game obtained by merging
the terminals in each half,
and then discard the irrelevant
half of the terminals (the one \emph{not} containing the value of the
starting vertex) and all vertices with the corresponding values. This method, which is due to
Mike Paterson, uses the optimal $O(n)$ comparisons, but requires
$\Theta(\log n)$ iterations, each with a worst case running time of $\Theta(m)$.
\subsubsection{$O(n)$ Comparisons and $O(m\log\log n)$ Time.}
To improve the running time of Paterson's algorithm, we stop and sort
the remaining terminals
as soon as this can be done in $O(n)$ time. The number of comparisons
is still $O(n)$.
As noted in Section \ref{prelim}, we may assume that all vertices have positive values.
\paragraph{Algorithm.}
Given an SRG $G$ with $m$ arcs, $n$ terminals, and starting position
$v_0$, do the following
for $i=0,1,2,\ldots$
\begin{enumerate}
\item \label{part} Partition the current set of $n_i$ terminals around
  their median payoff.
\item \label{solve} Solve the coarse game obtained by merging the
  terminals in each half.
\item Remove all vertices that do not have values in the half containing $\Val_G(v_0)$. 
\item Undo step \ref{part} for the half of $v_0$.
\label{end}
\end{enumerate}
When $n_i \log n_i \leq n$, stop and solve the remaining game by the
sorting method.
\paragraph{Analysis.}
Steps \ref{part}--\ref{end} can be performed in $O(m)$ time and $O(n_i)$
comparisons. The number of iterations is $O(\log n - \log f(n))$,
where $f(n)$ is the inverse of $n \mapsto n \log n$, and since this equals
$O(\log \log n)$ we have the
following.
\begin{theorem}
An SRG with $m$ arcs and $n$ terminals can be weakly solved in
$O(m\log\log n)$ time and $O(n)$ comparisons.\qed
\end{theorem}
\subsubsection{Almost-Linear Time.}
We can balance the partitioning and
retrograde analysis to achieve an almost linear running time, by a technique
similar to the one used in \cite{GabowTarjan:1988} and later generalized
in \cite{Punnen:1996}.\footnote{Note, however, that while the
  technique
is similar,
the problem of solving simple recursive games
does not
seem to fit
  into the framework of \cite{Punnen:1996}.}
Again, we assume that all vertices have positive values.

\paragraph{Algorithm.}
Given an SRG $G$ with $m$ arcs, $n$ terminals, and starting position
$v_0$, do the following
for $i=0,1,2,\ldots$
\begin{enumerate}
\item \label{part1} Partition the current set of $n_i$ terminals into
groups of size at most $n_i/2^{m/n_i}$.
\item \label{solve1} Solve the coarse game obtained by merging the
  terminals in each group.
\item Remove all vertices having values outside the group of $\Val_G(v_0)$. 
\item Undo step \ref{part1} for the group of $v_0$.
\label{end1}
\end{enumerate}
When $n_i/2^{m/n_i} < 1$, stop and solve the remaining game by the
sorting method.
\paragraph{Analysis.}
\label{logstar_analysis}
All steps can be performed in $O(m)$ time. For the first step we can do a  ``partial perfect quicksort'',
where we always partition around the median  and stop at
level $\lceil m/n_i \rceil +1$.

To bound the
number of iterations, we note that $n_i$ satisfies the recurrence
\begin{equation}
n_{i+1} \leq n_i/2^{m/n_i}\,,
\end{equation}
which by induction gives
\begin{equation}
n_i \leq \frac{n}{  b^{b^{b^{{\rotatebox{25}{\ldots}}^{\hspace{-0.06cm}b^{}}}}}}
\hspace{-1.05cm}\raisebox{-.29cm}{\rotatebox{-65}{\Bigg\}}\hspace{-.5cm}\raisebox{-.45cm}{$i$}}
\end{equation}
where $b = 2^{m/n}$.
Thus, the number of iterations is $O(\log^*_b n)$,
where $\log^*_b$ denotes the number of times we need to apply the base
$b$ logarithm function to get below $1$.
This is easily seen to be the same as 
$O(1 + \log^* m - \log^* \frac{m}{n})$. 
 We have now established the following.
\begin{theorem}
An SRG with $m$ arcs and $n$ terminals can be weakly solved in
$O(m + m (\log^* m - \log^* \frac{m}{n}))$ time.\qed
\end{theorem}

\begin{remark}
When $m = \Omega(n\log^{(k)} n)$ for some constant $k$, this bound is $O(m)$.
\end{remark}
\subsubsection*{Acknowledgements.}  We are indebted to Mike Paterson
for his contributions to this work. We would also like to thank Uri Zwick and
Gerth Brodal for helpful discussions.
%
%
%
%
%
\bibliographystyle{plain}
\bibliography{/users/koda/trold/artikler/game}
\end{document}